\newcommand{\beq}{\begin{equation}}
\newcommand{\eeq}{\end{equation}}
\newcommand{\bea}{\begin{eqnarray}}
\newcommand{\eea}{\end{eqnarray}}

\documentclass[prl,aps,twocolumn]{revtex4}
\usepackage{epsf}
\usepackage{epsfig}
\usepackage{graphics}
\begin{document}

\title{Unusual doping and temperature dependence of photoemission spectra from
manganites}

\author{Prabuddha Sanyal$^{1,3}$}
\author{Subhra Sen Gupta$^{2,3}$}
\author{Nandan Pakhira$^{3,4}$}
\author{\mbox{H. R. Krishnamurthy$^{3,4}$}}
\author{D. D. Sarma$^{3,4,5,6}$}
\author{T. V. Ramakrishnan$^{3,4,7}$}

\affiliation{$^{1}$Harishchandra Research Institute, Allahabad 211019}
\affiliation{$^{2}$S.N. Bose National Center for Basic Sciences, Kolkata 700098}
\affiliation{$^{3}$Centre for Condensed Matter Theory (CCMT),
Department of Physics, Indian Institute of Science, Bangalore
560012, India.}
\affiliation{$^{4}$Jawaharlal
Nehru Centre for Advanced Scientific Research, Bangalore 560064,
India.}
 \affiliation{$^{5}$Solid State and Structural
Chemistry Unit, Indian Institute of Science, Bangalore - 560012,
India.}
\affiliation{$^{6}$ Indian Association for Cultivation of Science, Kolkata 700032} 
\affiliation{$^{7}$Department of Physics, Banaras Hindu
University, Varanasi 221005, India.}


\begin{abstract}

A recent, major, puzzle in the core-level photoemission spectra of
doped manganites is the observation of a 1-2 eV wide shoulder with
intensity varying with temperature $T$ as the square of the
magnetization over a $T$ scale of order 200K, an order of magnitude
less than electronic energies. This is addressed and resolved here,
by extending a recently proposed two electron fluid $\ell-b$ model
for these systems to include core-hole effects. The shoulder arises
from a rapid redistribution of $e_g$ electron density, as a function
of $T$, between the highly localized($\ell$) and band-like ($b$)
states. Furthermore, our theory leads to a correspondence between
spectral changes due to increasing doping and decreasing $T$, as
experimentally observed.

\end{abstract}

\maketitle

Doped rare earth manganites, {\em R}$_{1-x}${\em A}$_{x}$MnO$_{3}$
({\em R} = rare earth ion; {\em A} = alkaline earth ion), show
exotic properties such as colossal
magneto-resistance~\cite{Jin-Chahara} (CMR) close to a temperature
($T$) driven Ferro-metal (FM) to para-insulator (PI) or para-metal
(PM) transition, an extraordinarily rich phase diagram with varying
$x$, $T$ and magnetic field ($H$), etc~\cite{mng-rev}. Particularly
intriguing among their properties is the redistribution of spectral
weight over an energy window of several $eV$ in
valence-band~\cite{DD,Saitoh} and Mn $2p$
core-level~\cite{Chainani-PRL,Tanaka} photoemission spectra (PES),
as a function of $x$ and of $T$. The changes in the spectral
features with doping, incompatible with even an approximately {\em
rigid-band} description, suggests qualitative and drastic
modifications of the underlying electronic structure with small
changes in $x$. Similarly, changes in $T$ affect the spectra over
energy ranges 100 times the thermal energy scale and cannot be
understood in usual terms. Curiously, changes in spectra with
increasing $x$ or with decreasing $T$ show striking
similarities~\cite{Saitoh}, suggesting a common mechanism. The
effects have been attributed~\cite{DD,Saitoh} to the unusually
strong electron-spin and electron-lattice couplings in these
compounds, but no specific theory has been proposed till now.

Conventional core-level photoemission calculations, in terms of
cluster models~\cite{cluster} or impurity models~\cite{impurity},
can not account for a drastic renormalization of the underlying
electronic structure with changing $x$ or $T$. The only known
mechanisms~\cite{Ohtaka} that yield $T$ dependent changes in PES are
Fermi-edge decoherence (FED) effects due to thermal excitations and
Debye-Waller type effects due to the scattering of electrons by
phonons. They lead to changes only on energy scales of order $T$ and
the Debye temperature respectively ($\sim 10$-$100~ meV$), and not
over several $eV$ as observed. Cluster
calculations~\cite{Chainani-PRL} with {\it adjustable fitting
parameters}, using an MnO$_{6}$ octahedron coupled to a single level
at the Fermi energy, reproduce the multiplet features of the
observed spectra; but no $x$ or $T$ dependence, of the magnitude
seen in the experiments, can arise unless the fitting parameters are
{\it artificially} allowed to vary with $x$ and $T$.

In this paper we adopt a radically different approach complementary
to the above schemes, in that we focus on the $x$ and $T$
dependence, while neglecting details of multiplet structure. We
employ the recently proposed {\it two-fluid `$\ell$-$b$'
model}~\cite{TVRPai1} in a dynamical mean-field theory
(DMFT)~\cite{dmft-rev} framework which successfully explains several
hitherto poorly understood low energy properties of doped
manganites. The active degrees of freedom in the manganites are the
twofold degenerate $e_g$ levels, the $t_{2g}$ core-spins of Mn, and
the Jahn-Teller (JT) optical phonon modes of the MnO$_{6}$
octahedra. There are three strong on-site interactions, {\it viz.}
the JT electron-phonon coupling which splits the two $e_g$ levels by
an energy $2E_{JT}$ ($\sim$ 0.5 - 1 $eV$), the ferromagnetic Hund's
coupling $J_{H}$ between the $t_{2g}$ and $e_{g}$ spins ($\sim$ 2
$eV$) and the $e_{g}$ electron Coulomb repulsion $U_{dd}$ ($\sim$ 5
- 7.5 $eV$)\cite{Bocquet}; all larger than the $e_{g}$ inter-site
hopping ($t$ $\sim$ 0.2-0.4 $eV$)~\cite{DD-Satpathy} The
`$\ell$-$b$' model\cite{TVRPai1} is an effective low energy
Hamiltonian which implicitly captures the crucial effects of these
interactions and the quantum dynamics of the JT phonons. It invokes
two types of $e_g$ electrons, one {\em polaronic} and {\em
localized} ($\ell$), and the other {\em band-like} and {\em mobile}
($b$), and is given by,
$$H_{\ell b}  =  (-E_{JT}-\mu)\sum_{i,\sigma} n_{\ell i\sigma} -\mu
\sum_{i,\sigma}n_{bi\sigma} $$
$$+ U_{dd} \sum_{i, \sigma}n_{\ell
i\sigma}n_{bi \sigma}  -  t\sum_{<ij>, \sigma}(b^{\dagger}_{i,
\sigma} b_{j,\sigma } + H.C.) $$
\begin{equation}
 - J_{H}\sum_{i}({\vec {\sigma}_{\ell
i}} + {\vec {\sigma}_{bi}}) \cdot {\vec{S}}_{i} -
J_{F}\sum_{<ij>}{\vec{S}_{i}} \cdot {\vec{S}_{j}}
\label{lbham}
\end{equation}
The  {\em polaronically trapped }  `$\ell$' species has site energy
$-E_{JT}$, and an exponentially reduced hopping ($\sim$ 1 $meV$)
which has been neglected, while the {\em non-polaronic} `$b$'
species (site energy $0$) has undiminished hopping $t \sim$ 0.2-0.4
$eV$. $J_{F}$ is a novel ferromagnetic {\em virtual double-exchange}
(VDE) coupling ($\sim 2~meV$) between the core spins, which arises
naturally in this model~\cite{TVRPai1}. The chemical potential,
$\mu$, imposes the doping determined filling constraint:
$\sum_{\sigma}(\langle n_{\ell \sigma}\rangle + \langle
n_{b\sigma}\rangle) = (1-x)$.

The `$\ell$-$b$' model is similar to the Falicov-Kimball model
\cite{fkm-rev}, and is exactly soluble in the DMFT~\cite{dmft-rev}
framework, with the $t_{2g}$ core spins ($\vec{S}_{i}$) being
approximated as classical vectors ($S\hat{\Omega}_{i}$) and $J_F$
treated in the Curie-Weiss mean field approximation~\cite{TVRPai1}.
The resulting {\em self-consistent impurity model}~\cite{dmft-rev},
for a specific site, at any temperature $T$ and in the limit of
$J_{H}\rightarrow\infty$, depends parametrically on $\Omega_{z}$
(the $z$ component of the unit vector representing the core-spin) at
that site~\cite{TVRPai1}. In order to calculate the Mn $2p$
core-level PES, we add to this a single `core-hole' level (labeled
`$c$') of positive energy $\epsilon_{c}\approx 647.6eV$  at the
impurity site~\cite{kim}. The core hole has an attractive Coulomb
interaction $U_{pd}$  ($=$  -6.5 $eV$) with both the `$\ell$' and
the `$b$' electrons. The resulting Hamiltonian reads~:
$$H_{CL}(\Omega_{z}) =
\sum_{i}(\epsilon_{i}-\mu)a^{\dagger}_{i}a_{i}+
\sum_{i}V_{i}(\Omega_{z})(a^{\dagger}_{i}b+b^{\dagger}a_{i})$$
$$-(E_{JT}+\mu)n_{\ell} +\tilde{J_{F}}\langle m\rangle \Omega_{z}
+({U}_{dd}n_{\ell}-\mu)b^{\dagger}b$$
\begin{equation}
+(\epsilon_{c}-\mu)n_{c}+U_{pd}n_{c}(b^{\dagger}b+n_{\ell})
\label{h-cl}.
\end{equation}
Here, the $a^{\dagger}_i$ s create {\em bath} electrons,
representing the `$b$' electrons at the other sites of the
lattice~\cite{dmft-rev}, with which the `$b$' electrons of the
chosen site hybridize; and which, for the purposes of this paper, we
have approximated as having a discrete grid of energies
$\{\epsilon_i\}$. The corresponding hybridization parameters
$V_{i}(\Omega_{z})$ are obtained from the $V(\epsilon; \Omega_{z})$
determined self-consistently in the DMFT~\cite{TVRPai1,dmft-rev}.
$\tilde J_{F}= 2zJ_{F}S^{2}$ ($z$ = coordination number) and
$\langle m\rangle$ is the magnetization.

Both $n_{\ell}$ and $n_{c}$ are conserved in $H_{CL}$ (Eq.
(\ref{h-cl})). Hence, its eigenstates can be separately calculated
in the 4 sectors: $(n_{\ell}, n_{c})$ $=(0,0)$, $(0, 1)$, $(1, 0)$
and $(1, 1)$, labeled here as $(I0)$, $(F0)$ $(I1)$ and $(F1)$,
respectively. One can therefore calculate a separate spectral
function $A_{cc}(\omega;n_{\ell},\Omega_{z})$ in each $n_{\ell}$
sector, and for each  $\Omega_{z}$, as~:
$$\sum_{m_0,m_1} \frac{e^{-\beta E_{m_0}}}{Z(\{m_{0}\})} |<m_1|m_0>|^{2}
\times\delta[E_{m_1}-E_{m_0}-\hbar \omega]$$ Here $m_{0}$, $m_{1}$
refer to the {\em many-body eigenstates} of $H_{CL}$ in the
$n_{c}=0$ and $n_{c}=1$ sectors respectively for the
$(n_{\ell},\Omega_{z})$ specified. The full core-hole spectrum is
given by the weighted average
$$ A_{cc}(\omega)= \sum_{n_{\ell}} \int_{-1}^{1}d\Omega_{z}W_{n_{\ell}}
(\Omega_{z})A_{cc}(\omega;n_{\ell},\Omega_{z})$$  where
$W_0(\Omega_{z})$ and $W_1(\Omega_{z})$ are statistical weights
obtained as
$W_{n_{\ell}}(\Omega_{z})$=$Z(n_{l},n_{c}$=$0,\Omega_{z})/Z$ using
$Z(n_{l},n_{c},\Omega_{z})$, the {\em constrained partition
functions}, calculated for $n_{c}$=0 and for particular values of
$n_{l}$ and $\Omega_{z}$; and $Z$ is the total partition function.

The Boltzmann factors in the expression for $A_{cc}(\omega)$ give
rise to the aforementioned FED effects in PES, which are very weak.
However, special to our model, and hence to manganites, are two
other, unconventional, sources of $T$ and $x$ dependence which are
much larger: first, the statistical weights $W_0$ and $W_1$ are
dependent on $x$ and $T$; second, as we show below, the spectra for
each ($n_{\ell}$,$\Omega_{z}$) themselves change with $x$ and $T$,
with redistribution of spectral weights over scales of $eV$, because
the self consistent hybridization parameters $V_i(\Omega_z)$ are
strongly $x$ and $T$ dependent~\cite{TVRPai1}. Hence, for
simplicity, we neglect the FED effects in this paper, by restricting
$m_0$ above to just the ground state in the appropriate sector.

We have calculated the spectra using two different methods: (1)
$H_{CL}$ is single-particle like in each of the sectors (I0), (I1),
(F0) and (F1), and can be {\em exactly diagonalized} for any
$\Omega_{z}$. The initial states, {\it i.e.} the {\em many-body
ground state} (GS) for each $n_{\ell}$, are obtained by filling up
the single-particle levels in the $I0$ and $I1$ sectors up to the
chemical potential~\cite{state-note}. The final ($n_{c}=1$)
many-body states are obtained by creating particle-hole ($p$-$h$)
excitations with respect to the corresponding ground states, in the
$F0$ and $F1$ sectors~\cite{state-note}. We find that the PES
spectrum is dominated by the {\em single} $p$-$h$ channel (spectral
weight $> 95 \% $) which is calculable~\cite{state-note} to very
high accuracy even for a dense grid for the {\em bath} electrons.
(2) More involved calculations including contributions from {\em
all} $p$-$h$ channels, but limited to using only 21 {\em bath}
states, have been carried out using the Lanczos recursive
algorithm~\cite{state-note}. In both cases, as is standard practice,
the discrete spectra obtained have been broadened using a Gaussian
broadening with $\sigma \sim 0.1~eV$. The {\em single} and {\em all}
$p$-$h$ spectra are practically identical, and the small missing
weight (3-5\%) in the {\em single} $p$-$h$ channel is visible only
on close inspection. All the calculations reported in this paper are
for model parameters appropriate for the LBMO thin-film samples of
Tanaka {\it et al}.~\cite{Tanaka}.

\begin{figure}[tbp]
\resizebox{8.5cm}{6cm}
{\includegraphics*[90pt,460pt][525pt,805pt]{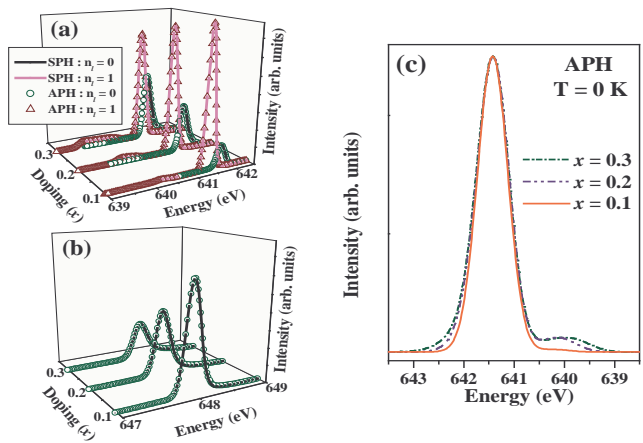}} \\
\caption{(Color online) (a),(b) Calculated core-level PES spectra at
$T=0$, shown in two parts for clarity, in the {\em single} $p$-$h$
(SPH) and {\em all} $p$-$h$ (APH) channels, for the $n_{l}=0$ and
$n_{l}=1$ sectors and for dopings 0.1, 0.2 and 0.3. (c) The fully
averaged APH spectra for the three dopings (normalized to the main
peak), showing only the main peak and the low energy shoulders.}
\label{doping}
\end{figure}

\begin{figure}[tbp]
\resizebox{8.5cm}{!}
{\includegraphics[100pt,515pt][500pt,790pt]{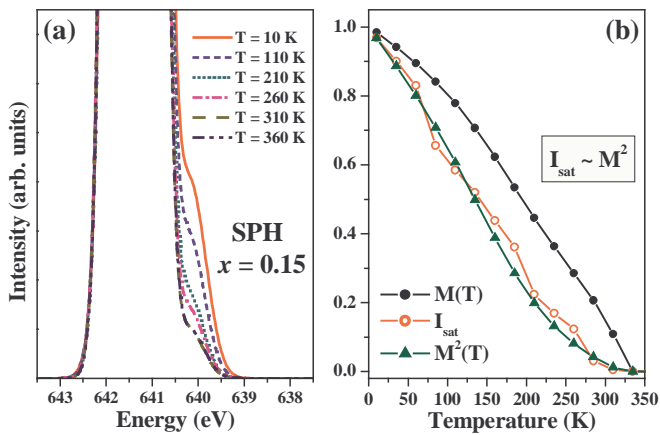}} \\
\caption{(Color online) (a) Magnified view of the $T$ dependence of
the calculated spectra averaged over the two $n_{l}$ sectors as well
as the core spin angles, in the {\em single} $p$-$h$ (SPH) channel
for $x$=0.15. (b) The $T$ dependence of the integrated shoulder
intensity in the difference spectrum with respect to the
paramagnetic spectrum, compared with $M^2(T)$, the square of the
(DMFT derived) magnetization. $M(T)$ is also shown.}
\label{temperature}
\end{figure}

Our results of the doping variation of the PES, obtained at $T=0$ in
the fully spin-polarised FM phase ($\Omega_z = 1$), are shown in
Fig. 1. Fig.s 1(a) and 1(b) show the spectra separately in the two
$n_{l}$ sectors and for both {\em single} and {\em all} $p$-$h$
channels, for $x = 0.1,~ 0.2,$ and $0.3$. The {\it fully averaged
spectrum} (Fig. 1(c)) thus has a {\em main peak} with {\em two
shoulders}, one on each side , separated by about 6.5 $eV$ ($\sim
|U_{pd}|$) from a high energy {\em correlation satellite} (not shown
in Fig. 1(c))~\cite{correl-sat}. As seen in Fig. 1(c), with
increasing $x$ till about 0.3, the main peak intensity decreases,
while those of the shoulders (and of the correlation satellite),
increase substantially~\cite{correl-sat}. This is similar to what is
seen in the core-level PES data in
LSMO~\cite{Chainani-PRL,correl-sat}. In the data on
LBMO~\cite{Tanaka} (where $x$ is not varied), only the low energy
shoulder is distinguishable. We believe that this is because LBMO
has a lower $x_{c}$ for the Ferro-Insulator (FI) to FM transition
($0.05$ as opposed to $0.16$ in LSMO), and hence~\cite{TVRPai1} a
smaller $(E_{JT}/D_{0})$. In that case the small $T$ dependent
changes in the higher energy shoulder, separated only by $\sim
E_{JT}$ from the main peak  which is rather broad in the experiment,
are harder to distinguish (see Fig. 2(a) where this effect is
simulated using a broadening of 0.3 eV).

Fig. 2(a) shows a magnified view of $T$ dependence of the fully
averaged spectrum in the {\em single} $p$-$h$ channel for $x$=0.15.
Clearly, spectral changes as {\it $T$ increases} are similar to the
changes as {\it $x$ decreases}, precisely as observed in
LSMO~\cite{Chainani-PRL}. In Fig. 2(b) we show the $T$ dependence of
the integrated shoulder intensity in the {\em difference spectrum}
with respect to the paramagnetic spectrum, and find that it tracks
$M^2(T)$, the square of the magnetization (as calculated using the
DMFT). Such a correspondence has indeed been observed experimentally
in LBMO~\cite{Tanaka}, and is not reproducible by conventional
rigid-band or cluster calculations.

All the major features of our calculated spectra can be understood
from the {\em single} $p$-$h$ channel contributions. Fig. 3 depicts
the initial state and two important types of final states in each of
the $n_{\ell}$ sectors in the metallic regime, with the bare
bandwidth $D_{0} = 1.3$ $eV$, $E_{JT}=0.29$ $eV$. The `$b$' band
occupancy ($n_{b}$) is small and the chemical potential, pinned at
$\mu \cong -E_{JT}$, lies close to the {\it effective} bottom edge,
$-\tilde{D}$, of the `$b$' band \cite{Deff}.

When $n_{\ell}=0$ (Fig. 3(a)) the local `$b$' level in the initial
state ($n_{c}=0$) is at zero and hybridizes sparingly with the
levels near the band edge, as the corresponding hybridization
amplitudes are small. Thus the local `$b$' character of the filled
levels in the initial state is small. When $n_c = 1$, the local
`$b$' level is pulled down by an amount $|U_{pd}| = $ 6.5 $eV$, and
becomes substantially occupied in the final GS. The local `$b$'
character of the occupied levels in the band is again very small.
Hence in the $n_{\ell}=0$ sector, the transitions involving $p$-$h$
excitations with energies close to and above the spectral {\em edge}
corresponding to the GS-to-GS transition
($\mu^{-}\rightarrow\mu^{+}$ in Fig. 3(a)), at an
energy~\cite{edge-com}  $\sim (U_{pd} + \epsilon_{c}-2\mu)\cong 641.68$
$eV$, have low intensity. The dominant contribution comes from the
transitions to final states where the local `$b$' electron is
excited to levels just above the chemical potential ($b\rightarrow
\mu^{+}$ in Fig. 3(a)). The corresponding {\em edge} is at an
energy~\cite{edge-com} $\sim (-E_{JT} + \epsilon_{c}-2\mu) \cong
647.89$ $eV$ and leads to the correlation satellite in Fig. 1(a).

\begin{figure}[tbp]
\resizebox{8.5cm}{11.5cm}
{\includegraphics[15pt,80pt][570pt,775pt]{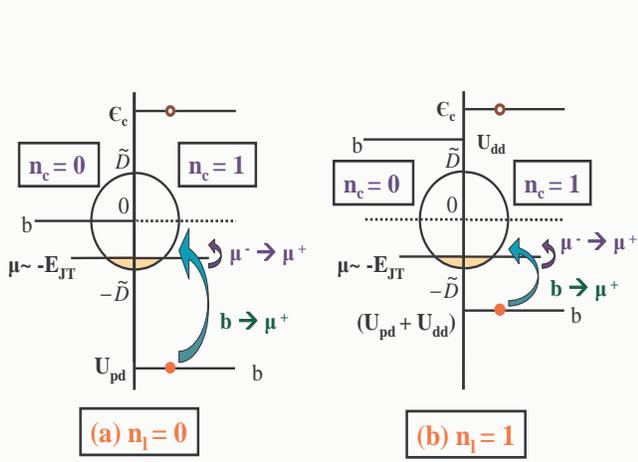}} \\
\vspace{-2.5 in} \caption{(Color online) Schematic depiction of the
GS configurations (with the occupied levels shown shaded) in the
initial state ($n_{c}=0$) and the ground and excited state
configurations in the final state ($n_{c}=1$) for the two sectors
$n_{l}=0$ (a) and $n_{l}=1$ (b). Transitions to final states with
the local `$b$' level occupied have little intensity. The dominant
transitions are to final states where the local `$b$' electron has
been transferred to just above the Fermi level.} \label{schematic}
\end{figure}
When $n_{\ell}=1$ (Fig. 3(b)), the local `$b$' level is pushed up,
by an amount  $U_{dd}=5$ $eV$, in the initial state ($n_{c} = 0$).
This further reduces the amount of `$b$' mixing and hence the local
`$b$' character of the occupied levels in the initial state. When
$n_{c} = 1$ (Fig. 3(b)), the `$b$' level is at  $(U_{dd} + U_{pd}) =
-1.5$ $eV$, much closer to the band edge than in the $n_{\ell}=0$
sector. Nevertheless, the GS when $n_c =1$ still has a substantial
occupancy of the local `$b$' level. Hence, the
$\mu^{-}\rightarrow\mu^{+}$ transitions (Fig. 3(b)) again have a
small intensity in the spectrum. The {\em edge} is now at an
energy~\cite{edge-com} $\sim (U_{dd}+2U_{pd} + \epsilon_{c} -
2\mu)=640.18$ $eV$,  1.5 $eV$ below the {\em edge} in the
$n_{\ell}=0$ sector. Just as in the $n_{\ell}=0$ sector, the main
contribution to the spectrum comes from the $b\rightarrow \mu^{+}$
transitions (Fig. 3(b)), beyond the {\em edge} at an
energy~\cite{edge-com} $\sim (U_{pd}-E_{JT} + \epsilon_{c} -
2\mu)=641.39$ $eV$, hence below the GS-to-GS {\em edge} from the
$n_{\ell}=0$ sector. Associated with the {\em edge spectra} are edge
singularities~\cite{singularity} and tails due to $p$-$h$
excitations, which, when smoothed out, give rise to the shoulders
with asymmetric~\cite{Doniach-Sunjic} lineshapes.

As one increases $x$ or decreases $T$, the `$b$' bandwidth increases
\cite{Deff}. However, $\mu$ still remains close to $-E_{JT}$, so
that the filling and the local `$b$' character of the occupied band
levels near $\mu$ increase in the initial state. Hence, one gets a
steady transfer of spectral weight from the features where the local
`$b$' level is unoccupied in the final state, to those where it is
occupied in the final state, as seen in the bare or un-averaged
spectra (Fig. 1(a)). The spectra shown in Fig. 1(b) are averaged
over the contributions from the two $n_{\ell}$ sectors (and
additionally over $\Omega_{z}$ for Figs. 2(a) and 2(b)), with
statistical weights which are themselves functions of $x$ and $T$.
The net effect is that the two shoulders on the two sides of the
main peak, arising from smoothed out edge spectra as shown above,
increase in intensity with increasing $x$ or decreasing $T$, in
agreement with experiments.

In conclusion, we have presented Mn $2p$ core-level PES calculations
by extending a new model for manganites \cite{TVRPai1} that takes
into account the simultaneous presence of strong {\em
electron-lattice}, {\em spin-spin} and {\em charge-charge}
interactions. Our results reproduce, for the first time, the unusual
redistribution of spectral weight over several $eV$ upon varying $x$
and $T$, and a correspondence between the effect of increasing $x$
and decreasing $T$, as experimentally
observed~\cite{Chainani-PRL,Tanaka}.

We would like to thank the JNCASR (NP,SSG) and the DST (HRK,PS,DDS)
for financial support.


\begin{thebibliography}{99}
\bibitem{Deff} As discussed in detail in \cite{TVRPai1},
the hopping of the `$b$' electrons is inhibited by an annealed
random medium of repulsive sites occupied by the `$\ell$' polarons.
At nonzero $T$,  the $t_{2g}$ spins (to which the `$\ell$' and `$b$'
spins are enslaved by the large $J_{H}$) get thermally disordered.
In either case, the resulting scattering leads to an effective
bandwidth $2\tilde{D}$ much smaller than the bare bandwidth $2D_0$,
but which increases substantially as $x$ increases or as $T$
decreases.
\bibitem{edge-com} The energies quoted in the text for the {\em edges}
are approximate as they do not
include the contributions from the hybridization terms.
\end{thebibliography}
\end{document}